\documentclass[
superscriptaddress,
pra,
twocolumn,
showpacs,
showkeys,
aps,
floatfix,
]{revtex4-1}

\usepackage{amsmath}
\usepackage{graphicx}
\usepackage{bm}
\usepackage{color}
\usepackage[colorlinks=true,linkcolor=blue,urlcolor=blue,citecolor=blue]{hyperref}

\newcommand{\rmc}{\mathrm{c}}
\newcommand{\rmd}{\mathrm{d}}
\newcommand{\rme}{\mathrm{e}}
\newcommand{\rmi}{\mathrm{i}}

\newcommand{\One}{\openone}
\newcommand{\eps}{\varepsilon}

\newcommand{\la}{\langle}
\newcommand{\ra}{\rangle}

\begin{document}
\title{Improving coherence with nested environments}

\author{H. J. Moreno}
\affiliation{Centro de Investigaci\'on en Ciencias, Universidad Aut\' onoma del Estado de
  Morelos,  Cuernavaca, Morelos, M\' exico.}
\affiliation{Instituto de Ciencias F\'\i sicas, Universidad Nacional
  Aut\' onoma de M\' exico, Cuernavaca, M\' exico.}

\author{T. Gorin}
\affiliation{Departamento de F\'\i sica, Universidad de Guadalajara,
   Guadalajara, Jal\'\i sco, M\' exico.}

\author{T. H. Seligman}
\affiliation{Instituto de Ciencias F\'\i sicas, Universidad Nacional
  Aut\' onoma de M\' exico, Cuernavaca, M\' exico.}
\affiliation{Centro Internacional de Ciencias A.C., Cuernavaca, M\' exico.}

\begin{abstract}
We have in mind a register of qubits for an quantum information 
system, and consider its decoherence in an idealized but typical situation. 
Spontaneous decay and other couplings to the far environment considered as the 
world outside the quantum apparatus will be neglected, while couplings to 
quantum states within the apparatus, i.e. to a near environment are assumed to 
dominate. Thus the central system couples to the near environment which in turn 
couples to a far environment. Considering that the dynamics in the near 
environment is not sufficiently well known or controllable, we shall use random 
matrix methods to obtain analytic results. We consider a simplified situation 
where the central system suffers weak dephasing from the near environment, 
which in turn is coupled randomly to the far environment. We find the 
anti-intuitive result that increasing the coupling between near and far 
environment actually protects the central qubit. 
\end{abstract}


\maketitle

In many quantum optics experiments and quantum information devices we find the 
following situation: The central system, well protected from simple decoherence 
processes such as spontaneous emission or direct coupling to a structureless 
heat bath, still suffers some decoherence from the coupling to the quantum part 
of the apparatus. We will call the former the far environment and the latter 
the near environment.
In such a tripartite system without direct coupling between central system and 
far environment, we find that increasing the coupling of the near to the far 
environment can protect the central system against decoherence. 
In the setting of the Haroche 
experiment~\cite{Bru96,Rai97} the late M. C. Nemes discussed this 
somewhat anti-intuitive fact
with one of the authors~\cite{nemesprivatecom} twelve years ago. More recently, 
additional numerical evidence has appeared 
~\footnote{C. Pineda, C. Gonzalez, and T. H. Seligman, (unpublished)}, some of 
which were master thesis related to the present 
work~\cite{Juan11, hector13, carlos14}. Finally, it was shown 
in Ref.~\cite{ZanCam14} that a protected subspace can appear in a strong 
coupling limit. 

We now wish to construct a model that allows some analytical treatment and 
simultaneously has some claim to universality. Indeed the intermediate 
environment in a quantum information system typically consists of quantum 
states that are not used but are unavoidably present. While ordered and 
systematic couplings can be minimized with good technology, it 
is plausible that the uncontrolled remanent will have a random structure. 
Considering its minimum information character~\cite{Bal68}, we use
random matrix theory (RMT) of decoherence~\cite{PGS07,GPKS08,CarrGorSel14}. 
There are several examples, which involve chaotic or irregular dynamics explicitly: 
{\it e.g.} the coupling of two-level 
atoms to quantum systems with classical chaotic analog~\cite{Haug05,WuToPr09}, 
and the experimental realization of quantum chaos in a chain of three level 
atoms in~\cite{Gra13}. The recent advances in the design and control of chains 
of individual ions~\cite{Pru11}, may lead to similar experimental systems.
We simplify the picture 
by limiting the coupling between central system and near environment to 
dephasing~\cite{GCZ97,GPSS04}, and assume the coupling between 
near and far environment to be of a tensor product form; see Eq.~(1), below. 
We can then take advantage of analytic expressions that exist for dephasing in 
absence of the far environment~\cite{GCZ97,GPSS04,GPS04,StoSch04b,StoSch05} and 
treat the effect of the far environment within a linear response 
calculation. We thus obtain analytic 
expressions for weak couplings between near and far environment, note that this 
range of coupling strengths is exactly opposite to that treated in 
Ref.~\cite{ZanCam14}. To study the effect of the far environment beyond the 
linear response approximation, we perform numerical simulations, using a 
modified Caldeira-Leggett master equation~\cite{CalLeg83}, whose equivalence to 
RMT models has first been discussed in Ref.~\cite{LutWei99}. 

{\it Model}: The full system consists of three parts, the central system, 
the near environment and the far environment with Hilbert spaces 
$\mathcal{H}_\rmc$, $\mathcal{H}_\rme$ and $\mathcal{H}_{\rm f}$, respectively. 
The unitary evolution of the entire system is given by the 
Hamiltonian
\begin{equation}
    H_{\rm tot} = H_0 + v_\rmc \otimes V_\rme \otimes \One_{\rm f}
    + \gamma\; \One_\rmc \otimes V'_\rme \otimes V_{\rm f}
\end{equation}
where $H_0 = h_\rmc \otimes \One_{\rme,{\rm f}} 
  + \One_\rmc \otimes H_\rme \otimes \One_{\rm f} 
  + \One_{\rmc,\rme} \otimes H_{\rm f}$.
Tracing out both environments leads to the reduced dynamics of the 
central system
$\varrho_c(t)= {\rm tr}_{\rme,{\rm f}}[\varrho_{\rm tot}(t)]$, with
\begin{equation}
\varrho_{\rm tot}(t)= \exp(-\rmi H_{\rm tot} t/\hbar)\; 
   \varrho_\rmc\otimes \varrho_{\rme,{\rm f}}\; 
   \exp(\rmi H_{\rm tot} t/\hbar) \; ,
\end{equation}
where $\varrho_\rmc$ and $\varrho_{\rme,{\rm f}}$ represent
the initial states of the central system (typically assumed to be pure), and 
the environment (near and far environment), respectively.
The couplings are given by the tensor products
$v_\rmc\otimes V_\rme$ (between central system and near environment) and
$\gamma\, V_\rme'\otimes V_{\rm f}$ (between near and far environment). The
former will be chosen as dephasing, such that $[h_\rmc,v_\rmc] = 0$. Such
couplings are frequently used, as they simplify calculations and maintain many 
essential properties.

{\it Dynamics}: We write the Hamiltonian as 
$H_{\rm tot} = \sum_j |j\ra\la j| \otimes\, H^{(j)}_{\rm e,f}$, with
\begin{equation}
H^{(j)}_{\rm e,f} = \big ( \eps_j\; \One_\rme + H_\rme 
   + \nu_j\; V_\rme \big ) \otimes \One_{\rm f} + \One_\rme \otimes H_{\rm f} 
   + \gamma\; V'_\rme \otimes V_{\rm f} \; ,
\end{equation}
where the set of states $\{\, |j\ra\, \}_j$ is a common eigenbasis of $h_\rmc$
and $v_\rmc$, while $\eps_j$ and $\nu_j$ are the corresponding eigenvalues. The
evolution of the whole system can be written as $\varrho_{\rm tot}(t) = 
\sum_{jk} \rho_{jk}(0)\, |j\ra\la k| \otimes\, \varrho^{(j,k)}(t)$, where
$\varrho_\rmc = \sum_{jk} \rho_{jk}(0)\, |j\ra\la k|$ is the initial state of
the central system, and
\begin{equation}
\varrho^{(j,k)}(t) = \exp\big (-\rmi H^{(j)}_{\rm e,f} t/\hbar\, \big )\;
\varrho_{\rme,{\rm f}}\; \exp\big (\rmi H^{(k)}_{\rm e,f} t/\hbar\, \big ) \; .
\end{equation}
We find for the matrix elements of the reduced state of the central system: 
$\rho_{jk}(t) = \rho_{jk}(0)\, {\rm tr}_{\rme,{\rm f}}[\varrho^{(j,k)}(t)]$.
Since ${\rm tr}_{\rme,{\rm f}}[\varrho^{(j,j)}(t)] = 1$, the diagonal elements
are constant in time, while the off-diagonal ones (i.e. the coherences) are
given as expectation values of generalized echo operators in the composite 
environment; see Eq.~(\ref{Def:relcoherence}) and Ref.~\cite{GPSZ06}. 

In other words focussing on an individual matrix element, $\rho_{jk}(t)$, we 
may introduce
\begin{equation}
H_\lambda = H_0 + \lambda\, V_{\rm eff} = H_\rme + \nu_j\; V_\rme \; , \, 
H_0 = H_\rme + \nu_k\; V_\rme \; ,
\label{HlambdaDecomp}\end{equation}
such that $\lambda\, V_{\rm eff} = (\nu_j - \nu_k)\, V_\rme$. This allows to 
connect the coherences for vanishing coupling to the far environment 
($\gamma\to 0$), with fidelity amplitudes~\cite{GCZ97,GPSS04}. Introducing the 
relative coherences
\begin{align}
f_{\lambda,\gamma}(t) &= 
   \frac{\rho_{jk}(t)}{\rho_{jk}(0)\, \rme^{-\rmi (\eps_j -\eps_k) t/\hbar}} 
\notag\\
 &= {\rm tr}_{\rme,{\rm f}}\big [ \rme^{-\rmi H_{\lambda,\gamma} t/\hbar}\;
   \varrho_{\rme,{\rm f}}\; \rme^{\rmi H_{0,\gamma} t/\hbar}\, \big ] \; ,
\label{Def:relcoherence}\end{align}
where $H_{\lambda,\gamma} = H_\lambda\otimes\One_{\rm f} 
   + \One_\rme\otimes H_{\rm f} + \gamma\; V_\rme'\otimes V_{\rm f}$,
we find that $f_{\lambda,0}(t) \equiv f_\lambda(t)$ with
\begin{equation}
f_\lambda(t)= {\rm tr}_\rme[M_\lambda(t)\, 
{\rm tr}_{\rm f}(\varrho_{\rme,{\rm f}})]\; , \quad
M_\lambda(t)= \rme^{\rmi H_0 t/\hbar}\, \rme^{-\rmi H_\lambda t/\hbar} \; .
\label{Def:tracefidamp}\end{equation}
Hence, $f_{\lambda,0}(t)$ becomes the fidelity amplitude for 
perturbing the Hamiltonian $H_0$ by $\lambda\, V_{\rm eff}$, given the initial
state ${\rm tr}_{\rm f}(\varrho_{\rme,{\rm f}})$ in the near environment. 

Modeling the effect of the far environment, we consider the simplest possible 
situation, where random matrix and master equation descriptions are 
equivalent~\cite{LutWei99}. This will allow to use a master 
equation for numerics and the random matrix model for analytics.
In Ref.~\cite{GPKS08}, it has been shown that without central system, the 
coherences in the near environment decay with the rate 
$\Gamma = 2\pi\, N_\rme\, \gamma^2/(\hbar\, d_{\rm f})$, which
is just $N_\rme$ times the Fermi-golden-rule rate for transitions between 
individual states. Here, $\gamma^2$ replaces the magnitude squared of the 
coupling matrix elements, since in our case $V_{\rm f}$ is chosen from a 
appropriately normalized random matrix ensemble, while $d_{\rm f}$ denotes the 
average level spacing (or inverse level density) of $H_{\rm f}$, and
$N_e$ is the dimension of the near environment. Choosing the 
Caldeira-Leggett master equation to describe the far environment, one obtains 
practically the same reduced dynamics. The only difference is that $\Gamma$ 
then depends on the dissipation constant and the 
temperature~\cite{Juan11,hector13}.

{\it Linear response calculation}: Applying the linear response approximation 
in the Fermi golden rule regime to the coupling between near and far 
environment (see appendix), one arrives at
\begin{equation}
    f_{\lambda,\Gamma}(t) \sim (1-\Gamma t)\; f_\lambda(t) + \Gamma
     \int_0^t\rmd\tau\; f_\lambda(\tau)\; f_\lambda(t-\tau) \; .
\label{P:mainresult}\end{equation}
The symbol $\sim$ means equal up to $\mathcal{O}(\Gamma^2)$, and from know on, 
we replace $\gamma$ by the physically more meaningful decoherence rate 
$\Gamma$. As we will see below, Eq.~(\ref{P:mainresult}) equation is valid as 
long as $\Gamma t \ll 1$. It constitutes our main result. The details of its 
derivation can be found in the appendix. Note that the result is valid for any 
Hamiltonian $H_\rme$, which may result in very different behaviors of 
$f_\lambda(t)$. The only necessary assumption is that $V'_\rme$ is sufficiently 
random.

Generally, we find that increasing the coupling strength to the far environment 
is indeed slowing down the decoherence in the central system. Depending on the
interaction strength between central system and near environment, and on the
functional form of $f_\lambda(t)$, the effect can be more or less pronounced. 
This can be demonstrated for generic systems, where 
one often finds that $f_\lambda(t)$ changes from an exponential decay in the 
Fermi golden rule regime to a Gaussian decay in the perturbative 
regime~\cite{CerTom02,GPS04}. In the former
the effect is zero, which can also be understood in physical terms. In that 
regime the temporal correlations 
$\la\tilde V_{\rm eff}(t) \tilde V_{\rm eff}(t')\ra$ of the perturbation in the 
interaction picture decay very fast -- on a time scale 
$t_{\rm corr} \ll t_{\rm dec}({\rm ce})$, the decoherence time in the central 
system. Therefore, even if the decoherence time in the near environment 
$t_{\rm dec}({\rm ef}) = \Gamma^{-1}$ is smaller than $t_{\rm dec}({\rm ce})$, 
as long as $t_{\rm dec}({\rm ef}) > t_{\rm corr}$, the far environment will 
have no effect on the decoherence in the central system.
In the perturbative regime by contrast, $f_\lambda(t)= \rme^{-\lambda^2 t^2}$,
such that $f_{\lambda,\Gamma}(t) \sim g_{\Gamma/\lambda}(\lambda t)$ with
\begin{equation}
g_\alpha(x)= (1 - \alpha\, x)\; \rme^{-x^2} + \alpha\; \sqrt{\pi/2}\; 
   \rme^{-x^2/2}\; {\rm erf}\big ( x/\sqrt{2}\big ) \; .
\end{equation} 
Although exact analytical results for $f_\lambda(t)$ 
exist~\cite{StoSch04b,StoSch05}, for simplicity, we will compare our results to 
the exponentiated linear response (ELR) expression from Ref.~\cite{GPS04}.
\begin{align}
f^{\rm ELR}_\lambda(t) &= \exp[-\lambda^2\, C(t)\, ] 
\label{ELRfidampres}\\
C(t) &= t^2 + \pi\, t - 4\pi^2\int_0^{t/(2\pi)}\rmd t'\int_0^{t'}\rmd t''\; 
      b_2(t'') \; ,
\notag\end{align}
where $b_2(t)$ is the two-point form factor~\cite{Meh2004}.

{\it Caldeira-Leggett master equation}:
For full-fledged numerical random matrix calculations, we would need to work in 
the Hilbert space of near and far environment. For the far environment, we 
would need a smaller mean level spacing in combination with a larger spectral 
span, as compared to the near environment. Still, in order to justify the use 
of RMT, we would need as many levels as possible also in the near environment. 
Such random matrix calculations are not viable, due to the dimension of the 
Hamiltonian matrices involved.

We will therefore use an approach which allows to work in the Hilbert space of 
the near environment alone, taking the effect of the far environment into 
account via the Caldeira-Leggett master equation~\cite{CalLeg83}, 
where we replace the diagonal
matrix representation of the harmonic oscillator Hamiltonian with a random
matrix, defined as in Eq.~(\ref{HlambdaDecomp}).
We choose both matrices $H_0$ and $V_{\rm eff}$ from 
the Gaussian orthogonal ensemble (GOE). We scale $H_0$ in such a way that the
mean level spacing becomes one in the center of the spectrum. The matrix 
elements of $V_{\rm eff}$ are chosen to have the variances 
$\la V^{\rm eff}_{ij}{}^2\ra = 1 + \delta_{ij}$. In that way, the strength of 
the perturbation (implied by the dephasing coupling to the central system),
measured in units of the mean level spacing $d_0$, is given by $\lambda$. In 
the following figures we scale time by the Heisenberg time 
$t_{\rm H}= 2\pi\hbar/d_0$. 

\begin{figure}
  \includegraphics[trim = 0.5mm 0mm 0mm 0mm , clip,width=0.5\textwidth]
  {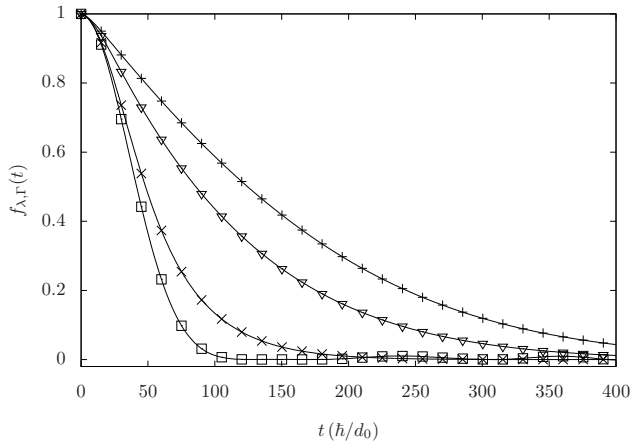}
\caption{Relative coherence $f_{\lambda,\Gamma}(t)$ in the Caldeira-Leggett 
model in the perturbative regime ($\lambda=0.02$), for different values of the 
coupling to the heat bath: $\Gamma/\lambda=0$ (squares), $1.0$ (shaped 
crosses), $5.0$ (inverted triangles) and $10.0$ (crosses). The time is in units 
of $\hbar/d_0$ where $2\pi\hbar/d_0$ is the Heisenberg time and $d_0$ the mean 
level spacing in the near environment. In all cases $N_\rme = 50$ and 
$n_{\rm run}= 1\, 000$ realizations. } 
\label{f:fidelity-l01}\end{figure}

{\it Numerical simulations}:
The use of random matrices requires a Monte Carlo average over many 
realizations. As a sufficiently large but still numerically manageable dimension
of the environment, we choose
$N_\rme = 50$, and perform averages over $n_{\rm run}= 1\, 000$ realisations. 
The general behaviour of the relative coherence $f_{\lambda,\Gamma}(t)$ is 
shown in Fig.~\ref{f:fidelity-l01}. Here, we choose $\lambda=0.02$
for the dephasing coupling,
and different values for the coupling strength $\Gamma$ between  
both environments. The figure clearly shows that the coherence decays slower
by increasing $\Gamma$.

\begin{figure}
  \includegraphics[trim = 10mm 0mm 0mm 0mm , clip,width=0.5\textwidth]
  {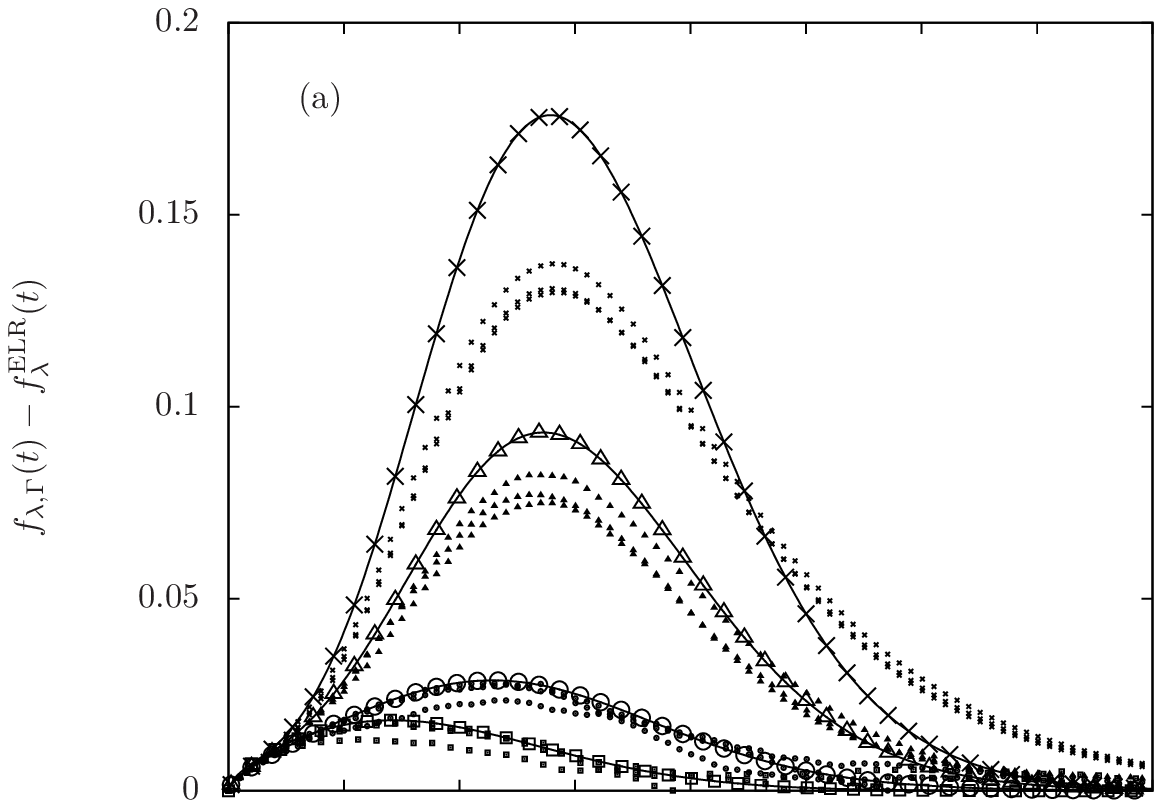}

  \vspace{-3mm}

  \includegraphics[trim = 10mm 0mm 0mm 0mm , clip,width=0.5\textwidth]
  {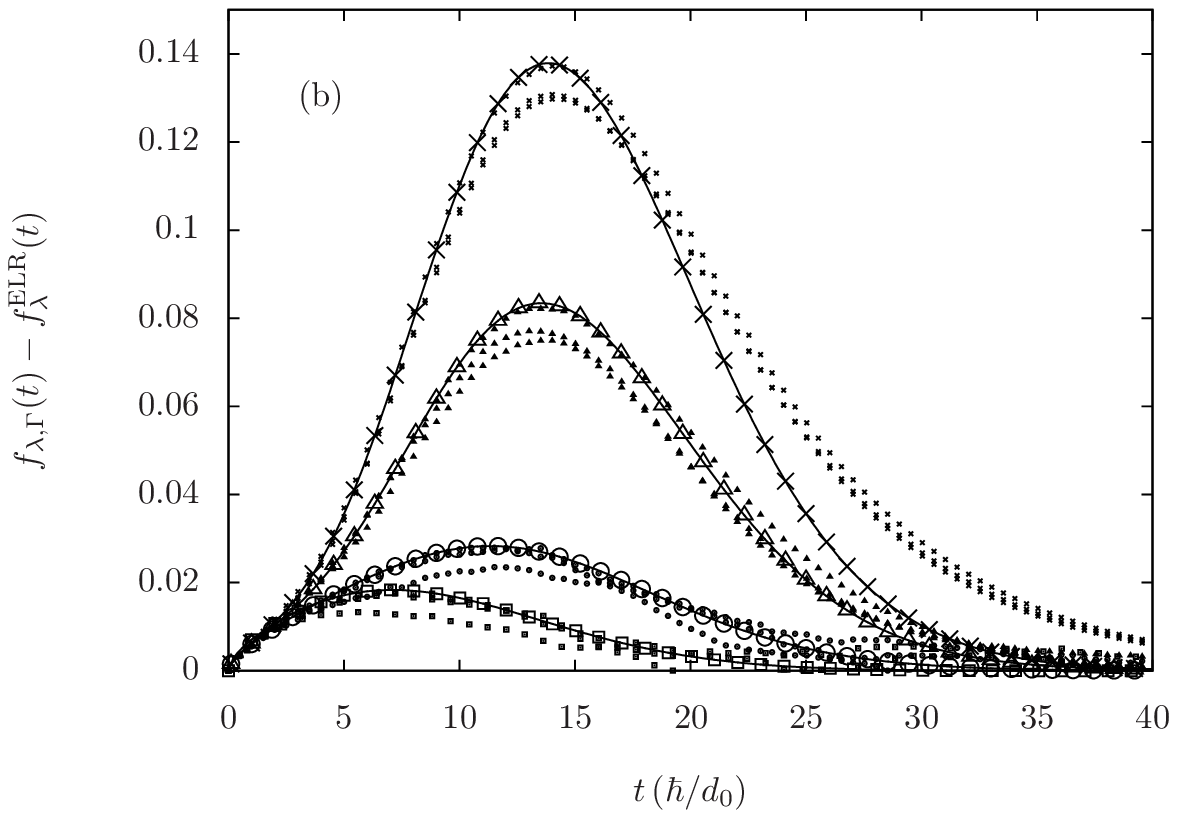}
\caption{(a) Relative coherence $f_{\lambda,\Gamma}(t)$ subtracted by the 
exponentiated linear response approximation $f^{\rm ELR}_\lambda(t)$ to the 
fidelity amplitude, for $\lambda = 0.1$ (cross-over regime) and different 
values for the coupling to the heat bath: $\Gamma/\lambda = 0$ (squares), $0.1$ 
(circles), $0.5$ (triangles), and $1.0$ (shaped crosses). The thick solid 
lines, show the analytical result according to Eq.~(\ref{P:mainresult}), and 
the nearest thin dotted lines show three statistically independent ensemble 
averages for each case. 
  (b) The same quantity as in panel (a), but for the theoretical curves the
values for $\Gamma$ are replaced by best fit values $\Gamma_{\rm fit}$ such that:    
$\Gamma_{\rm fit}/\lambda= 0.097$ (circles), $0.44$ (triangles), and $0.77$ 
(shaped crosses), obtained for the region $0 < t < 15$. The units of time
are the same as in Fig~\ref{f:fidelity-l01}.}
\label{f:triparty3}\end{figure}

In the remaining figures, we evaluate the quality of our 
analytical result from Eq.~(\ref{P:mainresult}). For a better quantitative 
comparison, we subtract the ELR approximation $f^{\rm ELR}_\lambda(t)$ for pure 
fidelity decay, from both, the numerical simulation and the analytical 
approximation for $f_{\lambda,\Gamma}(t)$. Note that for the function 
$f_\lambda(t)$ appearing in the analytical expression, we use numerical results 
with much improved accuracy.
These are obtained from numerical simulations without far
environment and some subsequent spline-fitting for facilitating the evaluation
of the integral in Eq.~(\ref{P:mainresult}). In that way, the accuracy can be
greatly improved.
The numerical result for $f_{\lambda,0}(t)$ differs from the ELR result due to 
the fact that $f^{\rm ELR}_\lambda(t)$ is only an approximation, but also 
because of the level density over the spectral range of $H_0$.
The remaining difference to the exact analytical expression found by 
St{\"o}ckmann and Sch{\"a}fer~\cite{StoSch04b,StoSch05}, is due to the fact 
that the trace in Eq.~(\ref{Def:tracefidamp}) includes the full spectral range 
where the level density varies according to the semi-circle law.

In Fig.~\ref{f:triparty3}(a) we consider the case $\lambda = 0.1$, 
which is in the cross-over regime. Since we are plotting the 
difference $f_{\lambda,\Gamma}(t) - f^{\rm ELR}_\lambda(t)$, the stabilizing 
effect of the far environment shows up as a growing positive hump. For each 
value of $\Gamma$, we plot three statistically independent numerical 
simulations. This gives us an idea about the statistical uncertainty of the 
results. We can clearly see that the curves
which correspond to $\Gamma = 0$ are different from zero, due to the reasons
discussed. Additional cases with increasing 
coupling to the heat bath. For those cases, the relative coupling strength
$\Gamma/\lambda$ is $0.1$ (circles), $0.5$ (triangles), and $1.0$ (shaped 
crosses).  We can observe that the theory agrees with the simulations, only in 
the case of smallest coupling, for stronger coupling the effect is 
systematically overestimated. In Fig.~\ref{f:triparty3}(b) we intend to 
find a rescaled decoherence rate $\Gamma_{\rm fit}$, 
which best describes the numerical results, and hence the stabilizing effect of
the far environment on the central system. A good agreement could be achieved 
only for times up to $t_{\rm max} \approx 15$, which is the approximate 
location of the maxima of the curves shown.
\begin{figure}
  \includegraphics[trim = 10mm 0mm 0mm 0mm , clip,width=0.5\textwidth]
  {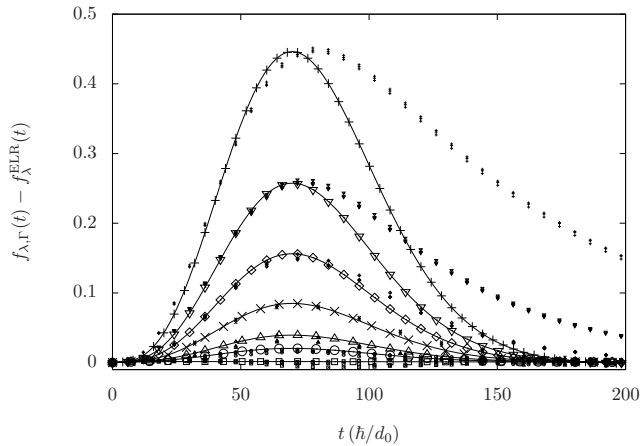}
\caption{Comparison between Caldeira-Leggett simulations and the linear 
response theory as in Fig.~\ref{f:triparty3}, but for $\lambda = 0.02$.
The units of time are the same as in Fig~\ref{f:fidelity-l01}.
For the theoretical curves, we rescaled $\Gamma_{\rm fit}$ as explained in the 
text. The different cases shown are: (squares) $\Gamma= 0$, circles 
$\Gamma (\Gamma_{\rm fit}) = 0.002$ ($0.00195$), triangles $0.004$ ($0.0039$), 
shaped crosses $0.01$ ($0.00858$), diamonds $0.02$ ($0.0159$), inverted 
triangles $0.04$ ($0.0263$), crosses $0.1$ ($0.0457$).}
\label{f:triparty5}\end{figure}

In Fig.~\ref{f:triparty5} we repeat the comparison for $\lambda = 0.02$, where 
the coupling between central system and RMT environment is close to the 
perturbative regime.  We use the same fitting procedure as in 
Fig.~\ref{f:triparty3}(b). Here, the values for the relative coupling strength
$\Gamma/\lambda$ range from $0.1$ to $5.0$. 
The slowing down of decoherence in the
central system due to the increasing coupling to the far environment, occurs
just as in the case $\lambda = 0.1$.
However, for large values of $\Gamma/\lambda$ the deviations between 
simulations and theory become quite noticeable, even if we use the best fit 
values $\Gamma_{\rm fit}$ for the theory.
\begin{figure}
  \includegraphics[trim = 5mm 0mm 0mm 0mm , clip,width=0.5\textwidth]
  {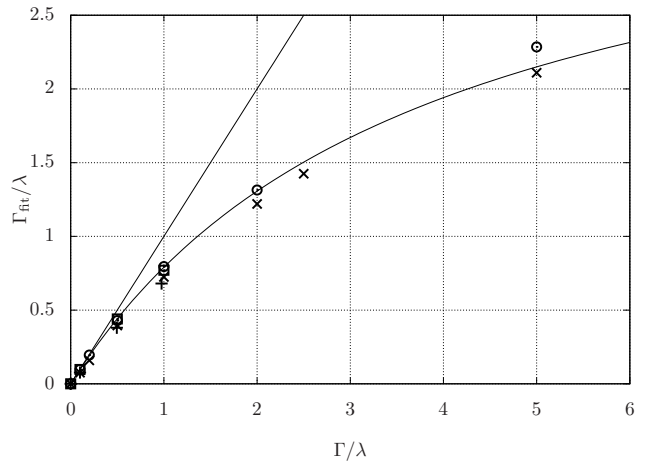}
\caption{$\Gamma_{\rm fit}/\lambda$ vs. $\Gamma/\lambda$ for 
$\lambda= 0.1$, $N_\rme= 25$ and $N_\rme= 50$ (crosses and squares 
respectively) and $\lambda= 0.02$, $N_\rme= 25$ (shaped crosses). The solid 
lines show the functions $\Gamma_{\rm fit}/\lambda = \alpha$ and 
$b\, \alpha/(b+\alpha)$ with $b= 3.77$ and $\alpha= \Gamma/\lambda$.}
\label{f:GfitvsGreal}\end{figure}

Finally, we compare in Fig.~\ref{f:GfitvsGreal} the fitted values 
$\Gamma_{\rm fit}$ for the coupling to the far environment, with the nominal 
ones, by plotting $\Gamma_{\rm fit}/\lambda$ versus $\Gamma/\lambda$. This is 
done for different dimensions of the near environment, for different coupling 
strengths between central system and near environment, and different couplings 
to the far environment. The derivation of our theoretical result within
linear response theory showed that the deviation from the exact result should
be quadratic in $\Gamma$. 
The results for $\Gamma_{\rm fit}/\lambda$ shown here, confirm this expectation,
as they approach the line $\Gamma_{\rm fit}/\lambda = \Gamma/\lambda$ for 
sufficiently small values.
For larger values of $\Gamma$, the fitted values for $\Gamma_{\rm fit}$,
and thereby the stabilizing effect of the far environment, increase ever more 
slowly.
To illustrate this behavior, we plotted the straight line 
$\Gamma_{\rm fit}/\lambda = \alpha$ as well as the function 
$g(\alpha)= b\, \alpha/(b+\alpha)$, with $\alpha= \Gamma/\lambda$ and a best 
fit value $b= 3.77$, which 
describes the overall behavior of the points quite well.

Summarizing, we have been able to obtain an analytic expression confirming
that nested environments can improve coherence of a central system as the 
coupling between near and far environment increases, as long as this coupling 
is small. We also extended previous limited numerical evidence for large 
coupling using a Caldeira-Leggett master equation which has been derived from 
RMT considerations in previous work~\cite{hector13}. This confirms that the 
effect subsists at large couplings between near and far environment, but 
subsides if the central system is strongly coupled to the near environment. 
An explanation on the basis of the quantum Zeno effect is tempting but 
problematic, at least in as far as we consider weak couplings between near and 
far environment.

\begin{acknowledgments}
We thank P. Zanardi, L. Campos Venuti, C. Gonzalez, and C. Pineda for 
enlightening discussions, and we acknowledge the hospitality of the Centro 
Internacional de Ciencias, UNAM, where many of these discussions took place. We 
also acknowledge financial support from CONACyT through the grants 
CB-2009/129309 and 154586 as well as UNAM/DGAPA/PAPIIT IG 101113.
\end{acknowledgments}

\begin{widetext}
\appendix*
\section{Derivation of the main result (Eq. 8)} 

\subsection{ Dephasing coupling}
Under dephasing coupling, the nondiagonal element of the qubit reduced state
is just the fidelity amplitude of the RMT-environment with respect to the
perturbation induced by the coupling between central system and near 
environment. For an initial state $\varrho_{\rm e,f}$, and with 
$V_{\rm e,f}= V_\rme'\otimes V_{\rm f}$ and 
$H_{\lambda} = H_0 + \lambda\, V_{\rm eff}$ from Eq.~(\ref{HlambdaDecomp}) of the
main article, 
 \begin{align}
     f_{\lambda,\gamma}(t) &= {\rm tr}\big [\, \varrho_{\rm e,f}\;
    \rme^{\rmi (H_0 + H_{\rm f} + \gamma\, V_{\rm e,f})t/\hbar}\; 
    \rme^{-\rmi (H_{\lambda} + H_{\rm f} + \gamma\, V_{\rm e,f})t/\hbar}\, 
 \big ] \notag\\
    &= {\rm tr}\big \{\; \big [\, 
    \rme^{\rmi (H_{\lambda} + H_{\rm f})t/\hbar}\; 
    \rme^{-\rmi (H_{\lambda} + H_{\rm f} + \gamma\, V_{\rm e,f})t/\hbar}\; 
    \varrho_{\rm e,f}\;
    \rme^{ \rmi (H_0 + H_{\rm f} + \gamma\, V_{\rm e,f})t/\hbar}\; 
    \rme^{-\rmi (H_0 + H_{\rm f})t/\hbar} \, \big ] \;
    \rme^{ \rmi (H_0 + H_{\rm f})t/\hbar}\; 
    \rme^{-\rmi (H_{\lambda} + H_{\rm f})t/\hbar}\; \} \; .
 \end{align}
The later two evolution operators are separable and therefore simplify as
follows:
 \begin{align} 
    \rme^{ \rmi (H_0 + H_{\rm f})t/\hbar}\; 
    \rme^{-\rmi (H_{\lambda} + H_{\rm f})t/\hbar}
    &= \rme^{ \rmi H_0\, t/\hbar}\otimes \rme^{ \rmi H_{\rm f}t/\hbar}\;\;
    \rme^{ -\rmi H_{\lambda} t/\hbar}\otimes \rme^{-\rmi H_{\rm f}t/\hbar} 
    \notag\\
    &= M_{\lambda}(t)\otimes\One_{\rm f}\; , \qquad 
    M_{\lambda}(t)= \rme^{ \rmi H_0\, t/\hbar}\; \rme^{ -\rmi H_{\lambda} t/\hbar} \; . 
    \notag
 \end{align}
Since ${\rm tr}_{\rm f}[ A\, M\otimes\One_{\rm f}] 
  = A_{ij,kl}\, M_{km}\delta_{lj} = {\rm tr}_{\rm f}(A)\, M$ then
\begin{equation}
    f_{\lambda,\gamma}(t) = {\rm tr}_{\rme}\big [ \, \tilde\varrho_\rme(t)\; M_{\lambda}(t)\, \big ], 
    \quad \tilde\varrho_\rme(t) =  {\rm tr}_{\rm f}\big [\,
    \rme^{\rmi (H_{\lambda} + H_{\rm f})t/\hbar}\; 
    \rme^{-\rmi (H_{\lambda} + H_{\rm f} + \gamma\, V_{\rm e,f})t/\hbar}\; 
    \varrho_{\rm e,f}\;
    \rme^{ \rmi (H_0 + H_{\rm f} + \gamma\, V_{\rm e,f})t/\hbar}\; 
    \rme^{-\rmi (H_0 + H_{\rm f})t/\hbar} \, \big ] \; .
 \label{D:fidgenform}\end{equation}

\subsection{ Linear response approximation for the coupling to the far
environment} 
The trace over the far environment in Eq.~(\ref{D:fidgenform}) is almost
exactly of the form as the reduced density matrix (in the interaction picture)
treated ~\cite{GPKS08}, 
namely with $\mathcal{M}_{\lambda,\gamma}(t)= \rme^{\rmi 
        (H_{\lambda} + H_{\rm f})t/\hbar}\; \rme^{-\rmi (H_{\lambda} + H_{\rm f} + \gamma
        V_{\rm e,f})t/\hbar}$,
we may write
 \begin{equation}
   f_{\lambda, \gamma}(t)= {\rm tr}_\rme\big [\, \tilde\varrho_\rme(t)\; M_{\lambda}(t)\, \big ] \; ,\qquad
     \tilde\varrho_\rme(t)= {\rm tr}_{\rm f}\big [\, \mathcal{M}_{\lambda,\gamma}(t)\; 
    \varrho_{e,f}\; \mathcal{M}_{0,\gamma}(t)^\dagger\, \big ] \; .
 \label{D:fidrhofarM}\end{equation}
However, in order to apply the formalism of~\cite{GPKS08}, 
we should assume the
coupling $V_{\rm e,f}$ and the initial state $\varrho_{\rm e,f}$ to be 
separable:
\[ \varrho_{\rm e,f} = \varrho_\rme\otimes \varrho_{\rm f} \; , \qquad
V_{\rm e,f} = V'_\rme\otimes V_{\rm f} \; , \] 
and: 
\[ \tilde V_{\rm e,f}(\tau) 
     = \tilde v_{\lambda}(\tau) \otimes \tilde V_{\rm f}(\tau)
     = \rme^{\rmi (H_{\lambda} + H_{\rm f})t/\hbar}\; V'_\rme\otimes 
          V_{\rm f}\; \rme^{-\rmi (H_{\lambda} + H_{\rm f})t/\hbar} \; , \qquad
      \tilde v_{\lambda}(\tau)= \rme^{\rmi\, H_{\lambda} t/\hbar}\; V'_\rme\; 
      \rme^{-\rmi\, H_{\lambda} t/\hbar} \; , \]
where we have already defined the representation $\tilde V_{\rm e,f}(t)$ of the 
coupling operator to the far environment in the interaction picture and 
similarly for $\tilde V_{\rm f}(\tau)$. Of course there remains the very 
important difference, that here we have different echo operators on the left 
and the right side of the initial state. Nevertheless, following carefully the 
calculation in~\cite{GPKS08}, 
developing the echo operators into their 
respective Dyson series we find:
 \begin{align}
   \tilde  \varrho_{e}(t) &= \varrho_\rme - \frac{\gamma^2}{\hbar^2} (A_J - A_I)\; , \qquad
     A_J = {\rm tr}_{\rm f}\big [\, J_{\lambda}(t)\, \varrho_{\rm e,f} 
     + \varrho_{\rm e,f}\; J_{0}(t)^\dagger \, \big ] \; , \quad
    A_I = {\rm tr}_{\rm f}\big [\, I_{\lambda}(t)\; \varrho_{\rm e,f}\;
    I_{0}(t)^\dagger \, \big ]
    \notag\\
    J_{\lambda}(t) &= \int_0^t\rmd\tau\int_0^\tau\rmd\tau'\; 
    \tilde V_{\rm e,f}(\tau)\, \tilde V_{\rm e,f}(\tau') 
    \; , \qquad
    I_{\lambda}(t)= \int_0^t\rmd\tau\; \tilde V_{\rm e,f}(\tau) \; .
 \end{align}
The calculation for the average over $J_{\lambda}(t)$ with respect to the 
random matrix $V_{\rm f}$ yields
 \begin{equation}
    \la J_{\lambda}(t)\ra= \int_0^t\rmd\tau\int_0^\tau\rmd\tau'\;
    c(\tau -\tau')\; \tilde v_{\lambda}(\tau)\; 
    \tilde v_{\lambda}(\tau') \otimes \One_{\rm f} \; ,
 \end{equation}
where $c(\tau)$ describes the spectral correlations of $H_{\rm f}$ and $\beta$
is the Dyson parameter, such that for a GUE ($\beta=2$) or a GOE ($\beta=1$)
with Heisenberg time $\tau_{\rm H}=2\pi\hbar/d_0$:
$c(\tau)= 3 - \beta + \delta(\tau/\tau_{\rm H}) - b_2(\tau/\tau_{\rm H})$.
Similarly for $A_I$:
 \begin{equation}
    \la A_I\ra = \iint_0^t\rmd\tau\rmd\tau'\; c(\tau -\tau')\; 
    \tilde v_{\lambda}(\tau)\; \varrho_\rme\; \tilde v_{0}(\tau') \; .
 \end{equation}
Finally, we obtain
 \begin{equation}
    \la A_J-A_I\ra = \int_0^t\rmd\tau\int_0^\tau\rmd\tau'\; c(\tau -\tau')\;
    \big \{\, \tilde v_{\lambda}(\tau)\; \big [\, \tilde
            v_{\lambda}(\tau')\,
    \varrho_\rme - \varrho_\rme\, \tilde v_{0}(\tau')\, \big ]
       - \big [\, \tilde v_{\lambda}(\tau')\, \varrho_\rme 
       - \varrho_\rme\, \tilde v_{0}(\tau')\, \big ]\; \tilde v_{0}(\tau)\, \big \}
 \label{D:LRfar}\end{equation}

\subsection{ Fermi golden rule regime and master equation}
If we assume that the Heisenberg time of the far environment $\tau_{\rm H}$ is
very large, and that we are in the Fermi golden rule regime for the coupling to
the far environment, then from the correlation function $c(\tau)$, we only need
to take the delta function into account. That reduces Eq.~(\ref{D:LRfar}) to
 \begin{equation}
    \la A_J-A_I\ra = \frac{\tau_H}{2}\int_0^t\rmd\tau\;
    \big [\, \tilde v_{\lambda}(\tau)\, \tilde v_{\lambda}(\tau)\, \varrho_\rme 
        -2\; \tilde v_{\lambda}(\tau)\, \varrho_\rme\, \tilde v_{0}(\tau)
    + \varrho_\rme\, \tilde v_{0}(\tau)\, \tilde v_{0}(\tau)\, \big ]
 \end{equation}
Next, we will average that expression over the coupling matrix $v_\rme$, which
is the near environment part of the coupling between near and far environment.
Since this matrix is assumed to be an element of the GUE, we find:
 \begin{equation}
    \la v_\rme^2\ra_{ik} = \sum_j v_{ij}\, v_{jk} = N_\rme\, \delta_{ik}
    \quad\Rightarrow\quad   
    \la \tilde v_{\lambda}(\tau)\, \tilde v_{\lambda}(\tau)\ra = N_\rme\; \One_\rme 
 \end{equation}
On the other hand, we find
 \begin{align}
     &\la\tilde v_{\lambda}(\tau)\, \varrho_\rme\, \tilde
     v_{0}(\tau)\ra_{iq}=
     (u^\dagger_{\lambda})_{ij}\, (v_\rme)_{jk}\, (u_{\lambda})_{kl}\, \varrho^\rme_{lm}\,
     (u^\dagger_{0})_{mn}\, (v_\rme)_{np}\, (u_{0})_{pq} \notag\\
        &\qquad = (u^\dagger_{\lambda})_{ij}\, \delta_{kn}\, \delta_{jp}\,
        (u_{\lambda})_{kl}\, 
        \varrho^\rme_{lm}\, (u^\dagger_{0})_{mn}\, (u_{0})_{pq}
        = (u^\dagger_{\lambda})_{ij}\, (u_{\lambda})_{kl}\, \varrho^\rme_{lm}\,
        (u^\dagger_{0})_{mk}\, (u_{0})_{jq} \; ,
 \end{align}
where $u_{\lambda}= {\rm exp}(-iH_{\lambda}t)$.
This can be written as
 \begin{equation}
     \la\tilde v_{\lambda}(\tau)\, \varrho_\rme\, \tilde
     v_{0}(\tau)\ra_{iq}=
     [M_{\lambda}(\tau)^\dagger]_{iq}\; {\rm tr}\big [\, \varrho_\rme\,
     M_{\lambda}(\tau)\, \big ]
    \quad\text{since}\quad
    M_{\lambda}(\tau)= u^\dagger_{0}\, u_{\lambda}\; . 
 \end{equation}
 Therefore, we obtain for $\varrho_{e}(t)$
 \begin{equation}
     \varrho_{e}(t)= \varrho_\rme - \frac{\gamma^2\, \tau_H}{2 \hbar^2}\left(
     2\, N_\rme\, t\; \varrho_\rme - 2\int_0^t\rmd\tau\;
 {\rm tr}\big [\, \varrho_\rme\, M_{\lambda}(\tau)\, \big ]\; M_{\lambda}(\tau)^\dagger \right)
 \label{D:tilrhoRes}\end{equation}
Let us denote $\Gamma = 2\pi \, N_\rme\, \gamma^2/(\hbar d_{\rm f})$, where we
introduced the average level spacing $d_{\rm f}= h/\tau_H$. Then we obtain for 
the fidelity amplitude:
\begin{equation}
    f_{\lambda,\Gamma}(t) = {\rm tr}\Big [\, (1 - \Gamma\, t)\; \varrho_\rme\;
        M_{\lambda}(t) 
         + \frac{\Gamma}{N_\rme}\int_0^t\rmd\tau\;
         {\rm tr}\big [\, \varrho_\rme\, M_{\lambda}(\tau)\, \big ]\;
         M_{\lambda}(\tau)^\dagger\;
     M_{\lambda}(t)\, \Big ] 
\label{D:fidMRes} \end{equation}
thus
\begin{equation}
f_{\lambda,\Gamma}(t) \sim (1 - \Gamma\, t)\; f_{\lambda}(t) + \Gamma\int_0^t\rmd\tau\;
    f_{\lambda}(\tau)\; f_{\lambda}(t-\tau)
 \label{D:fidLimRes}\end{equation}
 where we have used that $f_{\lambda}(t)= {\rm tr}[\varrho_\rme\, M_{\lambda}(\tau)]$ and
 $N_\rme\, f_{\lambda}(t-\tau) = {\rm tr}[M_{\lambda}(t-\tau)]$. So $f_{\lambda}(t)$ 
denotes the fidelity amplitude in the near environment, if there is no coupling 
to the far environment ($\gamma = 0$). The first line is exact (in the limit
$\Gamma t \ll 1$), assuming that no ensemble averaging has been applied with
respect to $H_\rme$ and $V_\rme$. The second line assumes self averaging for
the quantities 
${\rm tr}[\varrho_\rme\, M_\lambda(\tau)]$ and ${\rm tr}[M_\lambda(t-\tau)]$
which will probably hold for generic initial states $\varrho_\rme$ and
sufficiently large near environment ($N_\rme \gg 1$). 

\end{widetext}

\bibliography{JabRef}

\end{document}